# Cooperation sharing distributions in some cooperation-competition systems

X.-L. Xu, S.-R. Zou, C.-H. Fu, H. Chang, D.-R. He*

*College of Physics Science and Technology, Yangzhou University, Yangzhou 225002*

This online supplement mainly presents the details of empirical investigations of so-called "cooperation sharing" distributions in 12 real world cooperation-competition systems, which will be used for building an evolution model that will be reported elsewhere. All the distributions obey a function form $P(\omega) \sim (\omega + \alpha)^{-\gamma}$. The heterogeneities of the cooperation sharing distributions in different systems can be described by two distribution parameters, $\alpha$ and $\gamma$. The empirically determined system evolution durations are also presented.

## I. INTRODUCTION

Simultaneous cooperation-competition (C-C) of complex system elements have been a fascinating research topic in commercial [1-2], ecological [3-5], social [6-8], biochemical [8], economical [8-10] and some technological systems [11-18]. We proposed a model [18] in which elements simultaneously cooperate and compete in some groups. They unite together to create limited resources and compete for more "cooperation sharing" (CS). Matthew Effect generates a CS distribution, which interpolates between a power law and an exponential decay. In [18] we discussed the model analytically and then compared the analytic conclusion with the empirical results obtained in 12 real world C-C systems. However, since the statements on the details of the systems and their properties are long and boring, the information is suitable for this online supplement.

We are interested in a kind of C-C systems [11-18], which contain some C-C platforms (groups). The groups can be places, organizations, events, or activities. The system elements are C-C participants in the groups. Often in each group some elements make concerted effort to accomplish a task and create resources. When the elements cooperate, they also compete for a larger piece of the resources.

In the systems, we use $h_i$ to denote the group number in which element $i$ takes part. The quantity can be expressed as $h_i = \sum_j b_{ij}$ where $b_{ij}$ is defined as: $b_{ij} = 1$ if element $i$ joins group $j$ and $b_{ij} = 0$ otherwise. The so-called "group size", $T_j$, denotes the number of the elements, which take part in group $j$. $T_j$ is expressed as $T_j = \sum_i b_{ij}$ [11,12]. Fu et al. defined CS as the part of a countable resource, which the element shares [13]. However, in a more dissectional consideration, the competition intensity should depend on the group size. Therefore, if $Z_l$ denotes the total countable resource in group $l$, and $z_{il}$ denotes the part shared by element $i$, the CS of element $i$ in group $l$ was defined as $W_{li} = T_l z_{il}$. The normalized total cooperation sharing (NTCS), $\omega_i$, was defined as $\omega_i = (\sum_l T_l z_{il}) / \sum_j [(\sum_l T_l z_{jl})]$ [14,16,18].

In this supplement we present definitions and interpretations of 12 real world C-C systems. The NTCS distribution functions of the systems are also reported. All the distributions obey the so-called "shifted power law (SPL) functions" [12]. The function can be expressed as $P(x) \propto (x+\alpha)^{-\gamma}$. When $\alpha = 0$, it takes a power law form. In the condition that $x$ is normalized ($0 < x_i < 1$ and $\sum_{i=1}^{M} x_i = 1$, $M$ denotes the total number of $x$), we can prove [18] that SPL function tends to an exponential decay function when $\alpha \to 1$. Therefore an SPL interpolates between a power law and an exponential decay. The parameter $\alpha$ characterizes the degree of deviation from a power law. We emphasize that in general an SPL is not a power law, and $\gamma$ is not the power law scaling exponent. It is not strange to observe a very large $\gamma$ as will be discussed later. We shall show that both $\gamma$ and $\alpha$ characterize heterogeneity of the distribution, and that, very possibly, $\gamma$ and $\alpha$ keep a general correlation. When $\alpha \to 1$, $\gamma$ reaches very large values, indicating that the distribution function is near to an exponential decay.

In most cases, C-C between the elements occurs right after the birth of the groups; therefore, we can define "evolution duration" as the time duration between the group birth and the group termination or data collection. We define the "group birth" as the time when the first element joins it, and the "group termination" as the time when the cooperation task is accomplished and the elements disband.

The description of C-C dynamics certainly will be in a logarithmic time scale because the evolution durations of different systems must be very different and cross many orders of magnitudes. We emphasize that for the current purpose, it is sufficient that the data of the evolution duration are reliable within one order of magnitude, although the exact evolution duration for all the groups may be obtained. In some systems there are many groups, and the groups show different evolution durations. It is meaningless to list the evolution durations for all the groups. As listed in Table 1, we use the longest and the shortest evolution durations, $\tau_{max}$ and $\tau_{min}$, among the groups of the C-C system and the average of them in the current work, which are sufficient to capture the order of the magnitude and the error range of the evolution durations.

In section 2 we shall present empirical investigations. In section 3 a summary will be presented.

## II. EMPIRICAL INVESTIGATIONS

In constructing the real world C-C systems, the data are collected based on the following rules: (1) All the elements can be unambiguously assigned to the related groups; (2) The information of the amount of CS shared by the elements in each group is available; (3) The evolution duration of each group (this concept will be introduced bellow) is available. We totally constructed more than 60 systems, among which only 12 systems have all the reliable information. Therefore, the 12 systems can be considered to be randomly selected and reasonably cover the essential features of the C-C systems.

### A. 2004 Athens Olympic Game (OG)

In an Olympic game some athletes join a sport event to successfully conduct the pageant and also to obtain more sport scores. We define the athletes as the elements, the sport events (only the individual sport events, e.g., high jump, weight lifting, are considered) as groups and the sport scores as CS. The data were downloaded from www.sina.com.cn (2004), which includes 133 individual sport events and 4500 athletes, as well as their sport scores in each sport event. Figure 2(a) shows the empirical cumulative distribution of the NTCS, $P(\omega' \geq \omega)$. The fitting line parameters are $\alpha \cong 1$ and $\gamma=3617$. The corresponding fitting of exponential function is shown in Fig.2(b).

All the group evolution duration data can be downloaded. Within the groups, hectometer match shows the shortest act duration, which is 10.1 seconds. In comparison, Triathlon shows the longest act duration which is 6 hours and 25 seconds. The average of the act durations for this system is around 2 hours.

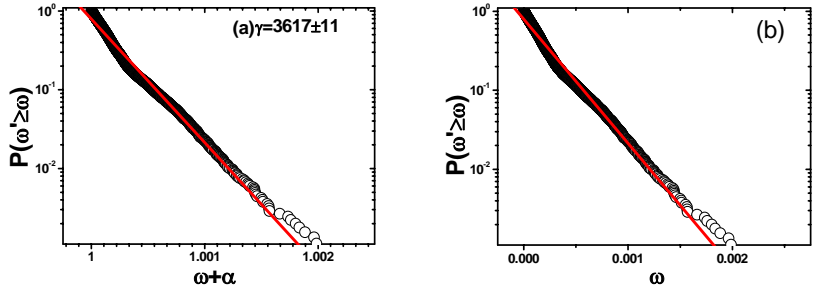

Fig. 1 (a). $P(\omega'\geq\omega)$ fitted by SPL function and (b). $P(\omega'\geq\omega)$ fitted by exponential function for the 2004 Athens Olympic Game.

**B. Chinese university matriculation system (CUM)**

In Chinese universities, colleges or departments are divided into undergraduate specializations. The undergraduates of different specializations are taught by different teaching schemes. In order to enter a specialization, a middle school student has to pass the national matriculation with the total marks higher than a lowest value. The university recruitment process is divided into several (roughly speaking, 6) batches in each province or region. Different batches have different lowest values of the matriculation mark. Different specializations may have the recruitment right in different batches depending on their academic levels. Also, each specialization may have the recruitment right in different batches in different geographical regions. A specialization with a higher batch can recruit higher mark students. Therefore, universities compete in the recruitment process in many "batches", which depend on specialization and geographical region, to recruit more and better students. From a different view point, all the universities are also cooperating in the total recruitment process to successfully complete the well-organized job. We define the batches as the groups, universities as the elements, and the lowest matriculation marks as CS, respectively. There are 51 batches and 2277 universities. The data were taken from the web stations www.hneeb.cn; www.jszs.net; www.lnzsks.com; and www.nm.zsks.cn. The NTCS cumulative distribution data, $P(\omega'\geq\omega)$, can be fitted by SPL functions (see Fig. 1(a)) with the parameters $\alpha\cong1$ and $\gamma$=2991. The data can be fitted by an exponential function as shown in Fig. 1(b).

Typically, the university matriculation lasts for several days during which the universities compete for better students. The lasting time durations are defined as the evolution durations. The data can be obtained from the above-mentioned web stations. For example, in 2006, the recruitment processes (from the time when the university delegacies get together to recruit students until the task is accomplished and they go home) of the first batch, the second batch, the third batch and the fourth batch in Shaanxi province lasted for 4 days, 4 days, 3 days and 4 days, respectively. Within all the batches, the longest act duration is 11 days, and the shortest one is 3 days. The average of the act durations for all the batches is around 6 days.

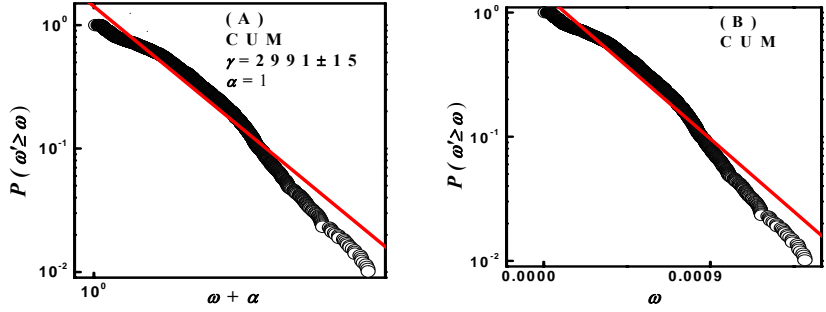

Fig. 2 (a). $P(\omega'\geq\omega)$ fitted by SPL function and (b). $P(\omega'\geq\omega)$ fitted by exponential function for the university matriculation system.

### C. Notebook PC selling at Taobao website (PCST)

In recent years, on-line shopping by internet becomes increasingly popular. Taobao (www.taobao.com) is the most famous on-line shopping mall in China where thousands of shops sell notebook PC. The shops collaborate to provide proper notebook PC selling service, and simultaneously compete for more profit. The shops are defined as elements, and a selling market of a notebook PC type is defined as a group. Usually, the price for the same type of notebook PC can be quite different in different shops. The shops with better reputation can sell out the same type of notebook PCs with higher price. Of course, these shops make more profit. Therefore, the selling price can be defined as CS of the element. Totally 53 notebook PC types and 4711 notebook PC shops from the Taobao on-line shopping mall were collected. Fig. 3 shows the cumulative distribution of the NTCS, $P(\omega'\geq\omega)$, and its SPL fitting. The parameters are $\alpha=0.00088$ and $\gamma=6.1$.

Refs. [19,20] provided the surveys, which let us know that, in average, a type of PC computer is updated every 3 years. Soon after the appearance of a new type of PC computer, the production of the old one stops. Accordingly, the C-C between the on-line shops for selling the old product terminates. Therefore, we believe that the reliable average act duration of a notebook PC product is 3 years, and the reliable logest and shortest act duration are 20 years and 0.5 year, respectively. This should be correct in order of magnitudes.

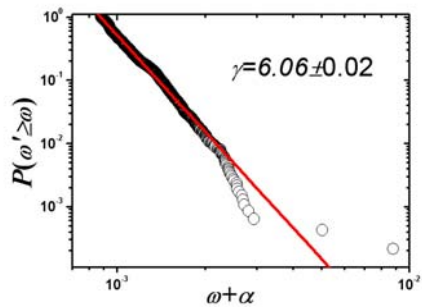

Fig. 3. $P(\omega'\geq\omega)$ for Taobao notebook PC selling system.

### D. Information technique product selling network (ITPS)

The market of information technique (IT) products, including mobile phone, computer, digital camera, and so on, is another example which shows typical C-C characteristics. If different manufacturers produce the same type of IT products, they compete in the selling market. Of course, these manufacturers also collaborate to supply enough IT products, and to maintain the market order. We define the manufactures as elements and the IT product selling markets as groups. On the website www.pcpop.com, there are detailed introductions to each IT product produced by a specific manufacturer. This web site also provides the "attention rank" data of the manufacturers for each IT product according to the total browsing time by the customers. To some extent, such "attention rank" is relevant to the competition abilities of the manufactures, and can be used to quantify the competition achievement, i.e., the profit. Therefore, we define the "attention rank" as CS. We collected 265 manufacturers and 2121 IT products from the website. Figure 4 shows the cumulative distribution of the NTCS, $P(\omega'\geq\omega)$, and its SPL fitting. The parameters are $\alpha=0.00135$ and $\gamma=4.465$.

It is difficult and pointless to determine the evolution duration for all the IT products. We randomly select 10 samples of IT products and find their exact evolution duration data. Then we count the shortest, longest and the average values of the durations within the samples, which are 94, 5 and 28 years, respectively (until 2006, the data collection time). We believe the order of magnitudes is reliable.

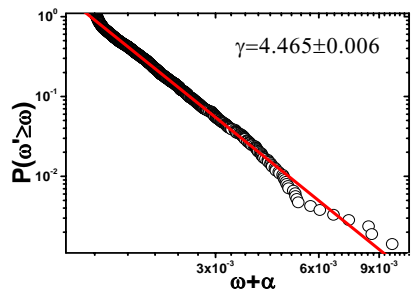

Fig. 4.  $P(\omega'\geq\omega)$ for IT product selling system

### E. Journal influence factor system (JIF)

Scientific journals are cooperating and competing in academic publication environment for higher scientific attention, which may be described by "influence factor" (IF) defined as the ratio of the total cited times of all the papers published in one year over the total paper number in the year. We define the journals as elements and the academic publication environment as a group. From website http://apps.isiknowledge.com/, we downloaded IF data of 6559 journals in 2006-2008. CS is defined as the averaged value of the IF data in the three years. Figure 5 shows the cumulative distribution of the NTCS, $P(\omega'\geq\omega)$, and its SPL fitting. The parameters are $\alpha=0.0022$ and $\gamma=3.88$.

Similarly we randomly select 10 sample journals and find their exact evolution duration data (at a website http://baike.baidu.com/), which is defined as the time durations from the time, when the journal originated, to 2006, the data collecting time. The journal with the longest duration was originated in 1812. The journal with the shortest duration was originated in 2005. Then we count the longest, shortest and the average values of the durations, which are 186, 3 and 34 years, respectively. We believe the order of magnitudes is reliable.

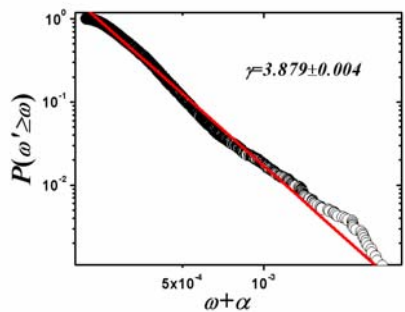

Fig. 5.   $P(\omega'\geq\omega)$ for Journal influence factor system

## F. Author academic level system (AAL)

We select the editorial board members of 15 journals with different IF values (from 34.48 (Nature in 2009) to 0.408 (International Journal of Modern Physics B in 2009)) as sample scientific paper authors. The journals are: (1) Nature, (2) Science, (3) Nature Photonics, (4) Nature Physics, (5) PNAS, (6) Physical Review Letters, (7) Physical Review E, (8) European Physics letters, (9) Chaos, Solitons and Fractals, (10) International Journal of Nonlinear Sciences and Numerical Simulation, (11) Physica A, (12) Chinese Physics B, (13) Chinese Physics Letters, (14) Modern Physics Letters B, (15) International Journal of Modern Physics B. The total number of the sample authors is 870. We define the scientists form a higher quality author collectivity. They play roles in scientific activities to maintain a proper order of scientific communication and development. This is a kind of cooperation. Simultaneously, as a scientific author, they compete in academic journal paper publication to achieve higher academic levels. Thus the sample authors are defined as elements, and the journal paper publication environment is defined as a group. The academic level, which is defined as CS, can be expressed as $\omega_i = \sum_j m_{ij} I_j$ where $m_{ij}$ denotes the paper number of author $i$ published in journal $j$ during 2006-2008, and $I_j$ denotes the averaged influence factor value of journal $j$ during the three years. We observed that, among the 870 authors, 86 ones did not publish papers in 2006-2008. We ignore these authors and present the empirical NTCS distribution result in Fig. 6. One can see that the data can be fitted by an SPL function with the parameter values $\alpha=0.0023$, $\gamma=3.65$.

Similarly we randomly select 10 sample authors and find their exact evolution duration data (at a website http://scholar.google.com/), which is defined as the time durations from the time, when the author published the first paper, to 2006, the data collecting time. We obtained that the author longest duration was $\tau_{max}$=61 years, and the shortest duration was $\tau_{minx}$=8 years. Then we count the average value of the durations as 29 years.

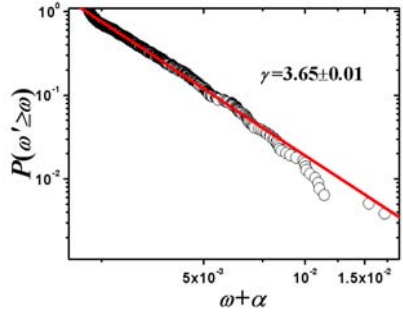

Fig. 6.   $P(\omega'\geq\omega)$ for Author academic level system

## G. The 200 richest Chinese magnates in 2004 (RCM)

We collected data about 200 richest Chinese magnates in 2004 from website http://www.forbes.com/lists/. In a view point, the magnates compete in the markets for a higher wealth rank. So we define the magnates as the elements and the general "market" as a group. Figure 7 shows that the CS, which is defined as the wealth ranks, distribution (NTCS distribution) obeys SPL function with the parameter values $\alpha=0.0012$ and $\gamma=3.1$.

From website http://baike.baidu.com/ we know the times when each magnate creates his career, or the creation time of the family enterprise. Then we know that the longest evolution duration is $\tau_{max}\approx100$ years, the shortest duration is $\tau_{minx}$=5 years, and the averaged duration is $\tau$=20 years.

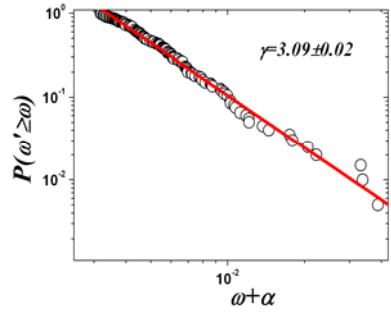

Fig. 7.  $P(\omega'\geq\omega)$ for 200 richest Chinese magnates system

## H.  The USA county population system in 1900 (UCO19)

We may think that USA counties compete in the region for more population. That is why some counties grow much faster than others. We defined counties as the elements and USA as a group. The population of each county is defined as its CS. We collected the data from http://www.census.gov./, including totally 2834 counties and $7.5778\times10^7$ people. Figure 8 shows the SPL NTCS distribution. The parameter values are $\alpha=0.00026$ and $\gamma=3.06$.

Due to difficulty for obtaining the exact time when each county was built, we collected the times when each county joined USA instead. Then we know that the longest evolution duration is $\tau_{max}\approx124$ years, the shortest duration is $\tau_{minx}=4$ years, and the averaged duration is $\tau=70$ years.

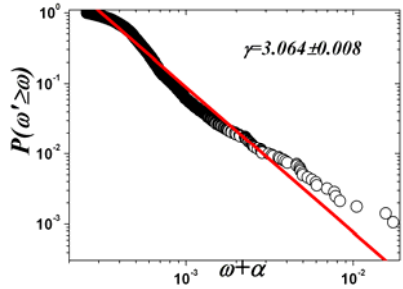

Fig. 8.  $P(\omega'\geq\omega)$ for USA county population (1900) system

## I.  The USA county population in 2000 (UCO20)

The interpretation and definitions of USA county population system in 2000 (UCO20) are similar as in the last subsection. We collected the data also from http://www.census.gov./, including totally 3142 counties and $2.8141\times10^8$ people. Figure 9 shows the SPL NTCS distribution. The parameter values are $\alpha=0.0001$ and $\gamma=2.2$. The longest evolution duration is $\tau_{max}\approx224$ years, the shortest duration is $\tau_{minx}=41$ years, and the averaged duration is $\tau=160$ years.

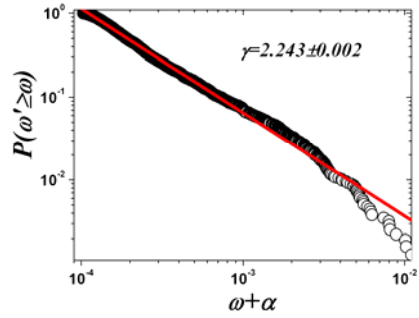

Fig. 9. $P(\omega'\geq\omega)$ for USA county population (2000) system

### J. Beijing restaurant system (BR)

Chinese food is famous in the world. Usually, one Chinese restaurant serves a large number of different cooked dishes, and the same cooked dish can be served by many restaurants. We define the restaurants in Beijing city as the elements and the cooked dish selling markets as the groups. In addition to collaborating to provide the food services, the restaurants serving the same type of dishes also compete to attract more customers and therefore earn more profits. The CS of each restaurant can be represented by the "attention degree" which is quantified by the customer marks given in the Dianping website (www.dianping.com/beijing) for each dish. We collected data of 688 cooked dishes in 3337 restaurants from the website until 2006. Fig.10 shows the cumulative distribution of the NTCS, $P(\omega'\geq\omega)$, of the Beijing restaurant network. The parameters are $\alpha=0.0000365$ and $\gamma=2.118$.

We randomly select 10 samples and find their act duration data. Then we count the shortest, longest and the average values of the act durations within the samples, which are 159, 9, 75 years until 2006, respectively. It should be reliable in the order of magnitudes.

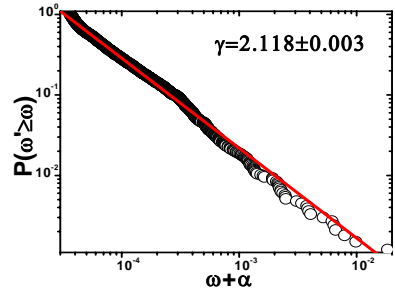

Fig.10. The $P(\omega'\geq\omega)$ of Beijing restaurants system

### K. Mixed drink (MD)

The mixed drinks (such as cocktails) usually contain a number of ingredients according to the consumer's taste, and many mixed drinks may share the same ingredients. We define the ingredients as the elements and the mixed drinks as groups. The ingredients collaborate to form mixed drinks with different tastes. Simultaneously, the ingredients contained in a common mixed drink compete since the ingredients have different relative importance. As the first step investigation, we very simply suppose that a certain ingredient in higher proportion is relatively more important. Therefore, we define the relative proportion of each ingredient in a certain mixed drink as the CS. Until 2006, we collected 7804 mixed drinks and 1501 ingredients. The proportions of the ingredients in each mixed drink are also obtained (www.drinknation.com). Fig.11 shows the cumulative distribution of the NTCS, $P(\omega'\geq\omega)$, of the mixed drink network and the SPL fitting. The parameters are

$α$=0.000051 and $γ$=1.783.

Similarly, we randomly select 10 samples and find their act duration data. Then we count the shortest, longest and the average values of the act durations within the samples, which are 500, 100, and 340 years, respectively. It should be reliable in the order of magnitudes.

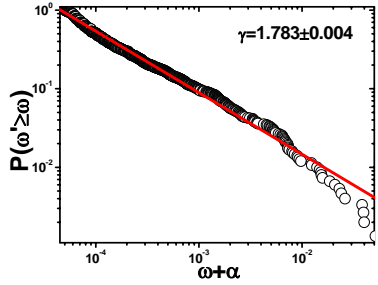

Fig.11. The $P(ω'≥ω)$ of mixed drink system

## L. The world language distribution system (WLD)

Based on the data from Ethnologue website (http://www.ethnologue.com, the 15[th] edition, published in 2005) which shows the populations speaking each language in a country (or a relatively independent region), we investigated the world language distribution system. The languages are defined as elements, and the countries are defined as groups. In a long time consideration, the languages compete to be used by more people. As the result, some languages have died out, but some other languages have been spoken by more and more people and spread to more and more geographical regions. The languages also collaborate in the common regions to accomplish the communications between the people. Therefore, we define the populations speaking a certain language in the country as CS. We collected 6142 kinds of languages used in 236 countries and regions. The total number of the language speakers is $5.2385×10^9$. Since the Ethnologue data are taken from multiple sources, the sum of the languages' population may not exactly equal to the total population in the world. Fig.12 shows the cumulative distribution of the NTCS, $P(ω'≥ω)$, for the word language network and its SPL fitting. The parameters are $α$=0.0000014 and $γ$=1.6316.

A group is defined as a country/region with its legal languages or the languages spoken by most of the people unchanged. If the languages basically change in the oral or written forms, we define that the old group dies out and a new group comes into the world. Chinese language ("Han" language, which has been used by most of Chinese people) has no basic change in about 5000 years although its oral and written forms have been developing in such a long period of time. As we know, this is the longest group evolution duration. One can mention quite some samples in which political or other factors induce the oral or written form basic change of the legal languages or the languages spoken by most of the people in a quite short time. For example, Singapore was governed by England for a long time and also by Japan and Malaysia for short time periods. However, the legal language was always English until 1965, when Singapore declared its autocephaly. The legal language changed to English, Chinese (Mandarin), Malay and Tamil since then. So, the group evolution duration of Singapore is 40 years (until 2005). As another example, in Vietnam, most of the people used Chinese as the written language for a long time, the presently used alphabetic writing language was created only 65 years ago during the France colonial domination. We believe that both the group evolution duration data (they are same in the order of magnitudes) represent the shortest group evolution duration.

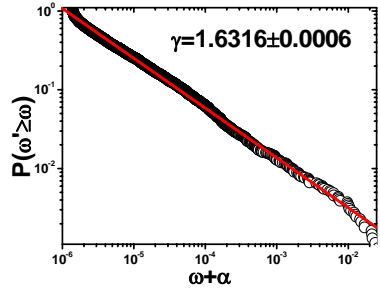

Fig.12.  The $P(\omega'\geq\omega)$ for the word language distribution system

We list the most important information in Table 1.

Table 1. The group and element interpretations and the numbers of them. Two parameter values, $\gamma$ and $N\alpha$ ($N$ denotes element number) of the NTCS distribution functions $P(x) \propto (x+\alpha)^{-\gamma}$. The longest and the shortest group evolution durations, $\tau_{max}$ and $\tau_{min}$, and the averaged evolution duration $\tau$.

| Syst. No. | system | group | element | CS | group No. | element No. | $\gamma$ | $N\alpha$ | $\tau_{max}$ (year) | $\tau_{min}$ (year) | $\tau$ (year) |
|---|---|---|---|---|---|---|---|---|---|---|---|
| 1 | OG | event | athlete | score | 133 | 4500 | $3.6\times10^3$ | $4.5\times10^3$ | $6.9\times10^{-4}$ | $3.2\times10^{-7}$ | $2.3\times10^{-4}$ |
| 2 | CUM | batch | University | mark | 51 | 2277 | $3.0\times10^3$ | $2.3\times10^3$ | 0.03 | 0.008 | 0.016 |
| 3 | PCST | market | shop | price | 53 | 4711 | 6.1 | 4.2 | 20 | 0.5 | 3 |
| 4 | ITPS | market | manufacturer | rank | 265 | 2121 | 4.5 | 2.9 | 94 | 5 | 28 |
| 5 | JIF | publication | journal | IF | 1 | 6559 | 3.9 | 1.4 | 186 | 3 | 34 |
| 6 | AAL | publication | author | level | 1 | 784 | 3.7 | 1.8 | 61 | 8 | 29 |
| 7 | RCM | market | magnate | wealth | 1 | 200 | 3.1 | 0.24 | 100 | 5 | 20 |
| 8 | UCO19 | USA | county | population | 1 | 2834 | 3.06 | 0.74 | 124 | 4 | 70 |
| 9 | UCO20 | USA | county | population | 1 | 3142 | 2.2 | 0.32 | 224 | 41 | 160 |
| 10 | BR | market | restaurant | attention | 688 | 3337 | 2.118 | 0.122 | 159 | 9 | 75 |
| 11 | MD | drink | ingredient | proportion | 7804 | 1501 | 1.783 | 0.0766 | 500 | 100 | 340 |
| 12 | WLD | country | language | population | 236 | 6142 | 1.6 | 0.0086 | $5\times10^3$ | 40 | $2.5\times10^3$ |

## III.  SUMMARY

We show in [18] an analytic discussion of a simple C-C evolution model where Matthew effect dominates so that different systems show different CS distribution heterogeneities only due to the different evolution durations. The analytically obtained relation functions between $\alpha$, $\gamma$ and the evolution duration, $\tau$, are in good agreement with the empirical results.

## ACKNOWLEDGMENTS

This work is supported by the National Natural Science Foundation of China under grant No. 10635040 and 70671089 and benefited of very helpful discussions with Dr. Shunguang Wu, Prof. Tao Zhou, Prof. Zong-Hua Liu and Prof. Hong Zhao.